\documentclass[aps,twocolumn,prl,superscriptaddress]{revtex4}

\usepackage[final,hiresbb]{graphicx}
\usepackage{amsmath,amssymb}

\usepackage{color,braket}
\usepackage{soul}

\begin{document}

\title{Identification and Resolution of Unphysical Multi-Electron Excitations in the Real-Time Time-Dependent Kohn-Sham Formulation}

\author{Xiaoning Zang}
\affiliation{Physical Sciences and Engineering Division (PSE), King Abdullah University of Science and Technology (KAUST), Thuwal, 23955-6900, Saudi Arabia}
\author{Udo Schwingenschl\"ogl}
\email{udo.schwingenschlogl@kaust.edu.sa}
\affiliation{Physical Sciences and Engineering Division (PSE), King Abdullah University of Science and Technology (KAUST), Thuwal, 23955-6900, Saudi Arabia}
\author{Mark T. Lusk}
\email{mlusk@mines.edu}
\affiliation{Department of Physics, Colorado School of Mines, Golden, CO 80401, USA}

\begin{abstract}

We resolve a fundamental issue associated with the conventional Kohn-Sham formulation of real-time time-dependent density functional theory. We show that unphysical multi-electron excitations, generated during time propagation of the Kohn-Sham equations due to fixation of the total number of Kohn-Sham orbitals and their occupations, result in incorrect electron density and, therefore, wrong predictions of physical properties. A new formulation is proposed in that the number of Kohn-Sham orbitals and their occupations are updated on the fly, the unphysical multi-electron excitations are removed, and the correct electron density is determined. The correctness of the new formulation is demonstrated by simulations of Rabi oscillation, as analytical results are available for comparison in the case of non-interacting electrons.
\end{abstract}

\maketitle

%%%%%%%%%%%%%%%%%%%%%%%%%%%%%%%%%%%%%%%%%%%%%%%%%
%\section{Introduction}
%%%%%%%%%%%%%%%%%%%%%%%%%%%%%%%%%%%%%%%%%%%%%%%%%

Time-dependent density functional theory (TD-DFT), the time-evolving analog to DFT, has moved into many facets of physics and chemistry since its formal foundation in 1984 with the Runge-Gross theorem \cite{RGtddftPRL1984}. Despite its ability to explicitly time-evolve many-body electronic states, the vast majority of TD-DFT applications is associated with excited state determination in response to weak electromagnetic fields. The external influence can then be treated as a perturbation and only the linear response of the system is needed to estimate excited states of ground-state systems. This kind of TD-DFT is referred to as linear-response (LR) TD-DFT \cite{TDDFT2} and has enjoyed many successes over the last few decades of service \cite{Ullrich2002, HeadGordon2004, Casida2009, Nakatsukasa2016}.

There are many physical settings, though, in which the external field is comparable to or even greater than the static electric field due to the nuclei or in which a system is driven continuously or starts from an excited state. Then the perturbative approach is no longer valid and time-explicit real time (RT) TD-DFT is appropriate. This approach is also useful, especially in nonlinear optical settings, in considering the evolution of excited states created from a laser pulse. Of course, a spectral decomposition of the results can always be carried out when the frequency-dependent properties are determined \cite{RTTDDFT1, Lopata2011}. For instance, RT-TD-DFT can obtain the first-order hyperpolarizabilities that characterize many second-order nonlinear processes, such as second harmonic generation and optical rectification \cite{Nonlinear2007}. It can also be used to calculate two-photon absorption cross sections \cite{Nonlinear2013}.

Aside from the consideration of nonlinear phenomena, the explicit time propagation of RT-TD-DFT makes it a promising computational tool for simulating many-body electron and exciton dynamics \cite{Lopata2011, Xiaosong2011, Cong2012, Yabana2012, Maitra2013, ZangTXPRA2017, ZangTXPRB2017, ZangSEPRB2017}. Its application becomes even broader when combined with molecular dynamics, as in, for instance, Ehrenfest-TD-DFT \cite{EMD2, EMD3, Xiaosong2007, FEMD, Xiaosong2009, EMD1}, which uses RT-TD-DFT for electron dynamics and classically describes nuclei motion via Ehrenfest molecular dynamics. Light-matter interactions are at the heart of RT-TD-DFT, being treated semi-classically. They may involve frozen nuclei, but it is possible to account for phonon interactions as well~\cite{Cong2012, Meng2010, Falke2014}. Either way, the electronic states are described in terms of linear combinations of ground-state Kohn-Sham (KS) orbitals.

In this work, we identify unphysical multi-electron excitations generated in the conventional formulation of RT-TD-DFT in the adiabatic approximation, which lead to wrong results in realistic systems. Rabi oscillation with a spin-independent Hamiltonian offers a particularly simple setting in which the problem is laid bare. It can be elicited in any atom, molecule, or nanostructure that has a sufficiently large interval between its lowest excited state and all other excited states to make a two-level approximation reasonable. Illumination with a resonant laser then results in the desired oscillation in ground and excited state occupations, i.e., the excited state occupation cycles between $0$ and $1$. The conventional formulation of RT-TD-DFT, though, exhibits a maximum excited state occupation of only $0.5$ (except for very specific excited states~\cite{FuksPRL2015}). One possible source of the Rabi occupation problem, scrutinized in the literature, is that the associated dipole moment is not properly captured. A different perspective was offered by Ruggenthaler \textit{et al.} who suggested that the problem has a classical origin \cite{BauerPRL2009}.  This view was later questioned by Fuks \textit{et al.} who instead attributed the issue to a lack of memory (history dependence) in the exchange-correlation functionals \cite{FuksPRB2011, FuksPRA2013}. More recently, Habenicht \textit{et al.} proposed two possible explanations: that there exists an artificial, computational multi-photon process that results in a three-level Rabi oscillation, or that the mean-field nature of DFT induces paired electron propagation \cite{BradleyJCP2014}.

The source of the problem is that the conventional formulation of RT-TD-DFT does not preclude a description of the excited states in terms of multi-electron excitations. While this may be physically reasonable under some circumstances, it is also possible that the formalism will offer a description that involves multi-electron excitations even though such terms are not present in the excited states predicted by LR-TD-DFT. We show that it is the unphysical multi-electron excitations that corrupts the electron density and track this back to a more fundamental problem: The fixed number of KS orbitals and their occupations. This problem can be eliminated using an update scheme in which the occupations are updated on the fly by calculating the so-called \emph{natural orbitals} \cite{Natural2010}. The formulation is therefore dubbed a time-dependent natural Kohn-Sham (TDNKS) formulation in contrast to the more traditional TDKS approach. Correctness of the proposed TDNKS formulation is demonstrated by simulations of Rabi oscillation.

%%%%%%%%%%%%%%%%%%%%%%%%%%%%%%%%%%%%%%%%%%%%%%%%
%\section{Time-Dependent Natural Kohn-Sham Formulation}
%%%%%%%%%%%%%%%%%%%%%%%%%%%%%%%%%%%%%%%%%%%%%%%%

% FIGURE #1
%
\begin{figure*}[ht]
\centerline{\includegraphics[width=1\textwidth]{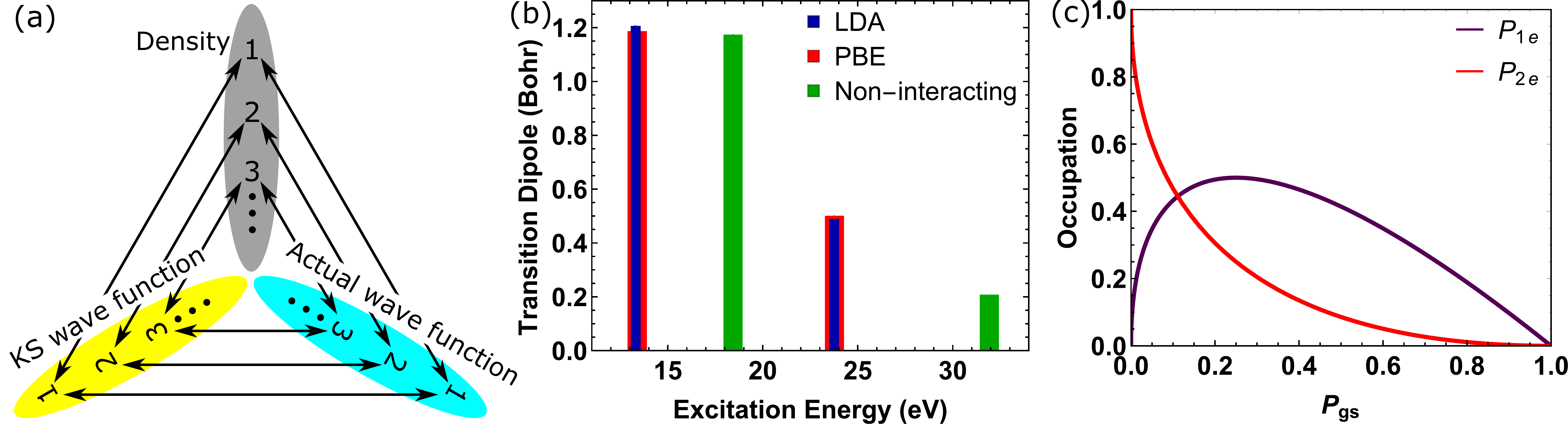}}\vspace*{-0.3cm}
\caption{(a) One-to-one correspondence between the KS and actual wave functions through the electron density (see also the Supplemental Material). (b) Transition dipoles from the ground state to the two lowest excited states of a ${\rm H_2}$ molecule for non-interacting and interacting (exchange-correlation functional treated in the local density approximation (LDA) or generalized gradient approximation parameterized by Perdew, Burke, and Ernzerhof (PBE)) electrons. (c) One-electron and two-electron excited state occupations as functions of the ground state occupation during Rabi oscillation. The maximum value of the one-electron excited state occupation is equal to $0.5$. }
\label{Population}
\end{figure*}

The KS formulation of TD-DFT leads to a set of single-particle equations coupled through density, 
\begin{eqnarray}
\mathrm{i}\hbar\frac{\partial}{\partial t}\psi_m(\vec r,t)&=&\Big[ -\frac{\hbar^2}{2m_e}\triangledown^2 +\nu_{ext}(\vec r,t)+\nu_{Hartree}[\rho](\vec r,t)\nonumber\\
&&+\nu_{xc}[\rho](\vec r,t)\Big]\psi_m(\vec r,t)\label{TD}\\
\rho(\vec r,t)&=&2\sum_{m=1}^{N}|\psi_m(\vec r,t)|^2 \label{dens_TDNKS}.
\end{eqnarray}
Here $\psi_m(\vec r,t)$ are the TDKS orbitals, while $\nu_{ext}$, $\nu_{Hartree}$, and $\nu_{xc}$ are the external, Hartree, and exchange-correlation potentials, respectively. Eq.\ (\ref{dens_TDNKS}) introduces the spin reduced electron density $\rho(\vec r,t)$ for $N$ KS orbitals and occupation $2$ for each KS orbital (to account for both spin orientations). The spin-restricted setting is natural in the sense that the Hamiltonian is spin independent. The direct consequence of this restriction is that the time-propagated multi-electron wave function is always a singlet, if the initial condition is a singlet. However, the fixed number of KS orbitals and fixed occupation of $2$ turn out to be the reason that full excited state occupation is not obtained during Rabi oscillation.

The TDNKS formulation is obtained by making a relatively simple change: The total number of orbitals involved and their occupations are updated as the system evolves. The spin-reduced density for our scheme is assumed to be
\begin{equation}
\rho(\vec r,t)=\sum_{m=1}^{N(t)}p_m(t)|\psi_m(\vec r,t)|^2 \label{newdens},
\end{equation}
where $N(t)$ implies that the number of orbitals is changing during time-propagation and  $p_m(t)$ are the evolving occupations. Note that the total number of electrons, and thus the summation of $p_m(t)$, is unchanged during time-propagation. Giving this reformulation upfront facilitates an explanation of what causes the problem with the traditional TDKS approach and how the TDNKS approach resolves the issue. 

The time-propagated orbitals, $\psi_m(\vec r,t)$, are represented as linear combinations of their initial forms, denoted as $\phi_n(\vec r)$:
\begin{equation}
\psi_m(\vec r,t)=\sum_na_{mn}(t)\phi_n(\vec r).
\label{KSexpand}\end{equation}
In the TDKS approach the time-propagated multi-electron wavefunction is a single determinant, $\ket{\Psi}=\ket{\psi_1(\vec r, t)\bar{\psi}_1(\vec r, t)\cdots\psi_N(\vec r, t)\bar{\psi}_N(\vec r, t)}$. During Rabi oscillation only two states should be involved: the ground state $\ket{\Phi_0}=\ket{\phi_1(\vec r)\bar{\phi}_1(\vec r)\cdots\phi_N(\vec r)\bar{\phi}_N(\vec r)}$ and one excited state. The excited state is a spin-adapted wavefunction, $\frac{1}{\sqrt{2}}(\ket{\Phi_i^r}+\ket{\bar\Phi_i^r})$, which is usually denoted as $^1\!\ket{\Phi_i^r}$ \cite{Szabo1989}. The notation implies that one electron is excited from the $i^{th}$ occupied orbital to the $r^{th}$ unoccupied orbital. The time-propagated multi-electron wavefunction must then be of the form $\ket{\Psi}=\ket{\phi_1(\vec r)\bar{\phi}_1(\vec r)\cdots\psi_i(\vec r, t)\bar{\psi}_i(\vec r, t)\cdots\phi_N(\vec r)\bar{\phi}_N(\vec r)}$, with $\psi_n(\vec r, t)=\phi_n(\vec r)$ and $\bar{\psi}_n(\vec r, t)=\bar{\phi}_n(\vec r)$ for $n\neq i$, and the time-propagated $i^{th}$ spin-up and spin-down orbitals are
\begin{eqnarray}
\psi_i(\vec r, t)&=&a_{ii}(t)\phi_i(\vec r)+a_{ir}(t)\phi_r(\vec r)\nonumber\\
\bar{\psi}_i(\vec r, t)&=&a_{ii}(t)\bar{\phi}_i(\vec r)+a_{ir}(t)\bar{\phi}_r(\vec r).
\label{TPorbital}\end{eqnarray}

We can re-write \cite{TimePropState}
\begin{eqnarray}
\ket{\Psi}&=&a_{ii}^2(t)\ket{\phi_i(\vec r)\bar{\phi}_i(\vec r)}\label{TPwfn} \nonumber\\
&+&\sqrt{2}a_{ii}(t)a_{ir}(t)\frac{\ket{\phi_i(\vec r)\bar{\phi}_r(\vec r)}-\ket{\bar{\phi}_i(\vec r)\phi_r(\vec r)}}{\sqrt{2}}\nonumber\\
&+&a_{ir}^2(t)\ket{\phi_r(\vec r)\bar{\phi}_r(\vec r)},
\end{eqnarray}
where we only show the time-propagated orbitals. For example, the ground state $\ket{\Phi_0}=\ket{\phi_1(\vec r)\bar{\phi}_1(\vec r)\cdots\phi_N(\vec r)\bar{\phi}_N(\vec r)}$ is simplified as $\ket{\phi_i(\vec r)\bar{\phi}_i(\vec r)}$ in this contracted representation. The second term of Eq.\ (\ref{TPwfn}) is the physical spin-adapted one-electron excitation. However there is also a third, unphysical two-electron excitation term, $a_{ir}^2(t)\ket{\phi_r(\vec r)\bar{\phi}_r(\vec r)}$, in which one spin-up and one spin-down electron are excited from the $i^{th}$ to the $r^{th}$ KS orbital. It is the consequence of spin restriction, i.e., the spin-up and spin-down states have the same time propagation according to Eq.\ (\ref{TPorbital}). Spin restriction makes all the terms in Eq.\ (\ref{TPwfn}) to singlets, as they should be, but this can also result in unphysical states. 

Two facts need to be clarified. First, the one-to-one correspondence between the KS and
actual wave functions established through the electron density, see Fig.\
\ref{Population}(a), ensures that we can study the occupations of the ground and excited
states by examining the KS wave function. Second, the lowest excited state of the
${\rm H_2}$ molecule does not include two-electron excitations according to Casida
perturbation TD-DFT. Note that the ${\rm H_2}$ molecule is well approximated by a
two-level system, since Fig.\ \ref{Population}(b) shows a large energy gap between the
lowest and second lowest excited states.

The unphysical third term of Eq.\ (\ref{TPwfn}) implies that the excitation is no longer that of a two-level system. We can write the one-electron excited state occupation, $P_{1e}$, and two-electron excited state occupation, $P_{2e}$, as functions of the ground state occupation, $P_{gs}$ (see also the Supplemental Material):
\begin{eqnarray}
P_{1e}&=&2\sqrt{P_{gs}}\left(1-\sqrt{P_{gs}}\right)\nonumber\\
P_{2e}&=&\left(1-\sqrt{P_{gs}}\right)^2\label{Popu}.
\end{eqnarray}
As shown in Fig.\ \ref{Population}(c), $P_{1e}$ can only reach $0.5$, consistent with the literature addressing the Rabi occupation problem. However, this incomplete one-electron excited state occupation is only one constituent of the total excited state occupation; the other piece is due to the unphysical two-electron excitation. The sum of these two excited state occupations and the ground state occupation is always equal to one.

Since the accompanying two-electron excitation is unphysical, we propose a reformulation in the construction of the multi-electron wavefunction of Eq.\ (\ref{TPwfn}) which results in only ground state and one-electron excited state contributions. The conventional KS formulation is based on an incorrect representation of the electron density in which unphysical two-electron excitation is the result of enforcing fermionic anti-symmetry and, at the same time, fixing the occupations. The consequence is incorrect occupation in Eq.\ (\ref{dens_TDNKS}). The new formulation, in terms of natural orbitals, is based on a correct electron density. Eq.\ (\ref{TPwfn}) can be recast as \cite{Szabo1989}
\begin{equation}
\ket{\Psi}=\sum_{m=i,r}\sum_{n=i,r}C_{mn}(t)\ket{\phi_m(\vec r)\bar{\phi}_n(\vec r)},
\label{recast}\end{equation}
where the matrix \textbf{C} gives rise to the electron density matrix, $\mathbf{\rho}=\mathbf{C}\mathbf{C^\dagger}$. After removal of the unphysical two-electron excitation $\ket{\phi_r(\vec r)\bar{\phi}_r(\vec r)}$ and renormalization of Eq.\ (\ref{recast}) the natural orbitals are obtained by direct diagonalization,
\begin{equation}
\mathbf{U^{\dagger}}\mathbf{\rho}\mathbf{U}=\mathbf{D}.
\end{equation}
Then $2\mathbf{D}_{mm}$ and $\sum_n\phi_n(\vec r)U_{nm}(t)$ are the new occupations, $p_m(t)$, and natural orbitals, $\psi_m(\vec r,t)$, of Eq.\ (\ref{newdens}), respectively. The new multi-electron wavefunction can be rewritten in natural orbitals as
\begin{equation}
\ket{\Psi}=\sum_{m=i,r}\sum_{n=i,r} B_{mn}\ket{\psi_m(\vec r,t)\bar{\psi}_n(\vec r,t)}\label{newTPwfn}
\end{equation}
with $\mathbf{B}=\mathbf{U^*}\mathbf{C}\mathbf{U^\dagger}$. The number of electrons is conserved, $\sum_{m=1}^{N(t)} p_m(t)=2N(0)$, because of the renormalization, and the constructed electron density is N-representable, since the multi-electron wave function of Eq.\ (\ref{newTPwfn}) is still antisymmetric. The energy is conserved during the simulation (see the Supplemental Material for further details). The name of our new scheme, TDNKS, is motivated by the fact that $\{\psi_m(\vec r,t)\}$ are the natural orbitals of Ref.\ \cite{Natural2010}. The procedure of a TDNKS simulation is summarized in Fig.\ S1. Although the multi-electron KS wave function and the one-electron reduced density are involved to obtain the correct electron density, the TDNKS approach is just a reformulation of TD-DFT, i.e., the equations of motion are fundamentally different from those of time-dependent reduced density matrix functional theory~\cite{Marques2012}.

%%%%%%%%%%%%%%%%%%%%%%%%%%%%%%%%%%%%%%%%%%%%%%%%%%
%\section{Computational Results}
%%%%%%%%%%%%%%%%%%%%%%%%%%%%%%%%%%%%%%%%%%%%%%%%%%

Rabi oscillation on an ${\rm H_2}$ molecule is used to compare the results of TDKS and TDNKS for predicting both the occupations and oscillation period. A  sinusoidal electric field is applied. If its energy is nearly resonant with the first excited state of the molecule, a Rabi oscillation will ensue. For the ${\rm H_2}$ molecule, the first excited state must be a linear combination of one-electron determinants, because the symmetries of one-electron and two-electron determinants are ungerade and gerade, respectively, so that the matrix elements between them are zero \cite{Szabo1989}. As a result, physically realizable eigenstates will not be a mix of the two. On the other hand, both spin-up and spin-down electrons feel the same external potential, so they will evolve in step with one another. This implies that at least part of the wavefunction will be due to two-electron excitations (with gerade symmetry) in the TDKS formulation. In addition to being unphysical, the nonzero occupation of the two-electron excited state implies an incomplete occupation of the one-electron excited state in the TDKS formulation. 

The real-space RT-TD-DFT implementation in Octopus \cite{OCTOPUS} is used. The simulation box is constructed by adding spheres around each atom with a radius of 3 \AA\ each, and  the space is discretized in increments of 0.15 \AA. Troullier-Martins pseudopotentials are employed and the LDA or PBE exchange-correlation functional is used. The simulation time step is set to be $\Delta t=4.84\times 10^{-19}$ s. The number of natural orbitals $N(t)$ and their occupations $p_m(t)$ are updated every $5\Delta t$, because the accuracy is not improved by smaller steps and the computation cost is reduced. For the bond length, the numerically optimized value of 0.74 \AA\ is used.

\begin{figure}[t]
\centerline{\includegraphics[width=0.5\textwidth]{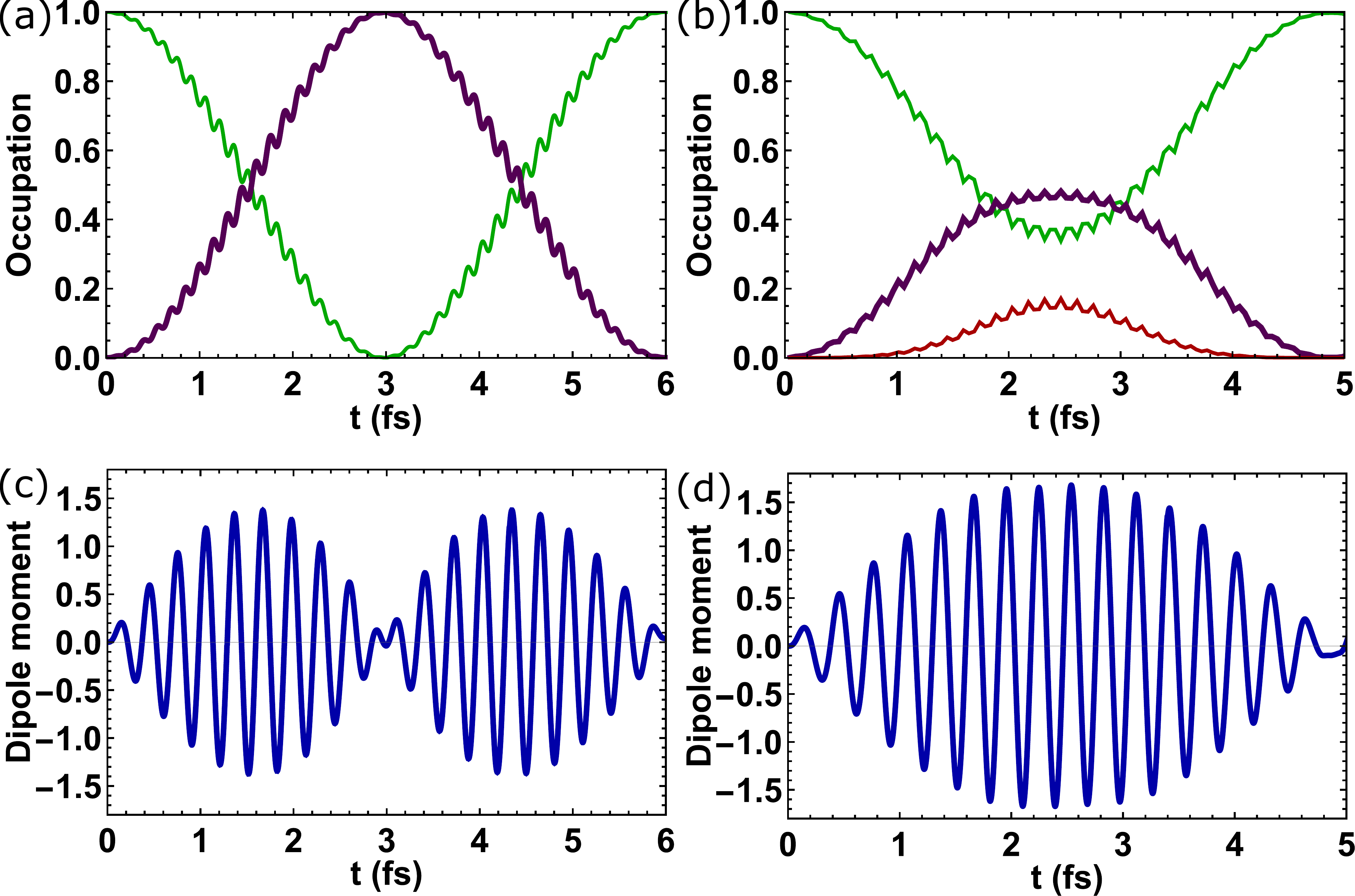}}\vspace*{-0.3cm}
\caption{Resonant Rabi oscillation from the (a) TDNKS and (b) TDKS approaches. The occupations of the ground state, $P_{gs}$ (green), one-electron excited state, $P_{1e}$ (purple), and two-electron excited state, $P_{2e}$ (red), are shown. Corresponding dipole moments are given in (c) and (d).}
\label{Rabi}
\end{figure}

Light-matter interaction is accounted for using the electric dipole approximation,
\begin{equation}
H_1=\vec{d}\cdot\vec{E}_0\cos(\omega t),
\label{H1_TDNKS}\end{equation}
where $\vec{d}$ is the transition dipole from ground state to first excited state and $|\vec{E}_0|=1.03$ V/\AA\ is the amplitude of the applied laser light. As there are only two electrons, the ground state is $\ket{\phi_1(\vec r)\bar{\phi}_1(\vec r)}$ and the one-electron excited state is $\ket{\Phi_1^2}=(\ket{\phi_i(\vec r)\bar{\phi}_r(\vec r)}-\ket{\bar{\phi}_i(\vec r)\phi_r(\vec r)})/\sqrt{2}$. As shown in Fig.\ \ref{Rabi}, the TDKS approach can only generate a maximum value of the one-electron excited state occupation of $0.5$ and there is always an accompanying two-electron excited state occupation. On the other hand, the TDNKS approach successfully generates full Rabi oscillation between the ground state and first excited state. We stress that it is justified to study the KS wave function instead of the actual wave function because of their one-to-one correspondence. Without realizing this fact, in former TDKS works about the Rabi oscillation the dipole moment was calculated to conclude about the occupation. According to time-dependent configuration interaction theory, for resonant excitation the dipole moment should be maximal when both the ground and excited states have an occupation of $0.5$ and minimal when either of them has zero occupation~\cite{FuksPRB2011}. As shown in Figs.\ \ref{Rabi}(c, d), these features are captured by the TDNKS but not by the TDKS approach. We calculate the dipole moment as trace of the product of the electron density and position operators, such that the correctness of the electron density is indicated by the correctness of the dipole moment. Since the electron density is correct, all other physical properties can be calculated as in the TDKS approach.

% FIGURE #3
%

\begin{figure}[t]
\centerline{\includegraphics[width=0.5\textwidth]{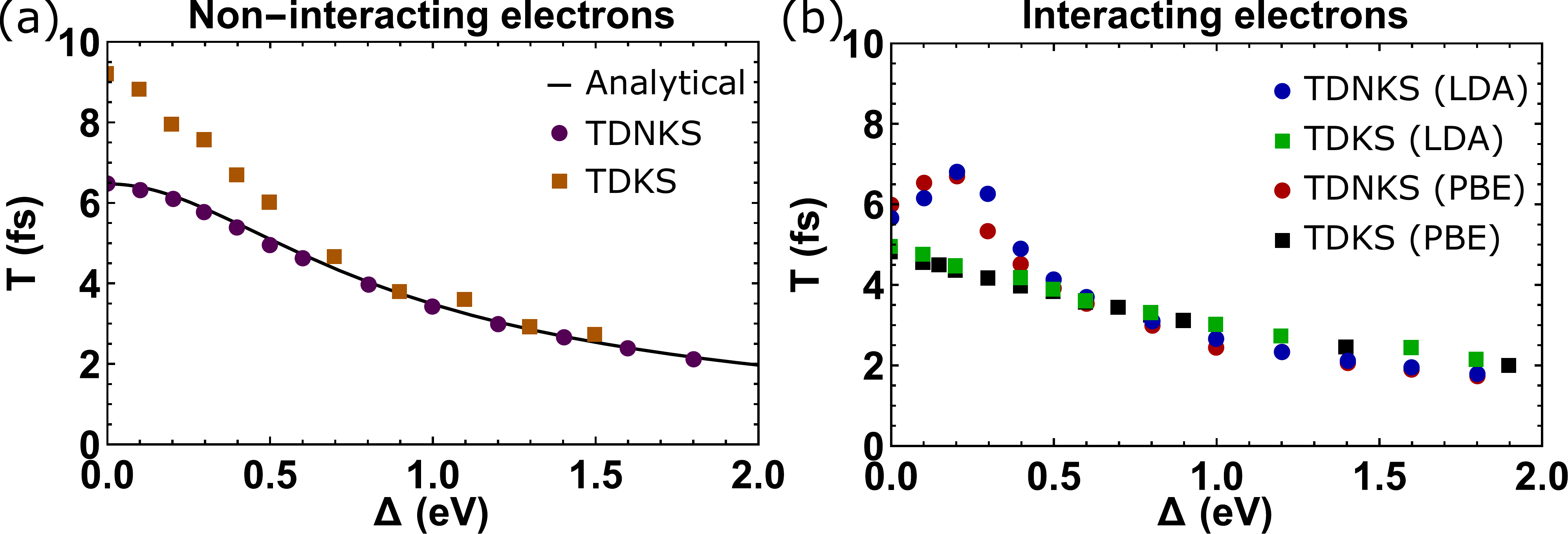}}\vspace*{-0.3cm}
\caption{Rabi oscillation period versus detuning energy. (a) The analytical result (from Eq.\ (\ref{period}); black line) is compared with numerical results from the TDNKS and TDKS approaches for non-interacting electrons. (b) Analogous numerical results for interacting electrons (no analytical solution exists).}
\label{Detune}
\end{figure}

In order to further test the accuracy of the TDNKS formulation, the Rabi oscillation period~\cite{Band2006}
\begin{equation}
T=\left[\left(\frac{\vec{d}\cdot\vec{E}_0}{2\pi \hbar}\right)^2+\Delta^2\right]^{-1/2}
\label{period}\end{equation}
is calculated as a function of the detuning energy $\Delta$, see Fig.\ \ref{Detune}. For the transition dipole we obtain from Casida perturbation TD-DFT amplitudes of $1.17\ {\rm Bohr}$ for non-interacting electrons and $1.21\ {\rm Bohr}$ (LDA), $1.18\ {\rm Bohr}$ (PBE) for interacting electrons. In these three cases a resonant Rabi oscillation is generated by laser light with energy $18.4\ {\rm eV}$, $13.3\ {\rm eV}$, and $13.4\ {\rm eV}$, respectively, see Fig.\ \ref{Population}(b). Fig.\ \ref{Detune}(a) shows the Rabi oscillation periods predicted by the TDKS and TDNKS approaches with the Hartree and exchange-correlation potentials turned off in order to model the ${\rm H_2}$ molecule with non-interacting electrons. The non-interacting electrons do not influence each other while they evolve with time. Thus, the Rabi oscillation period as a function of the detuning energy is expected to be given by Eq.\ (\ref{period}), which is the case for the TDNKS approach but not for the TDKS approach according to Fig.\ \ref{Detune}(a). Note the perfect match between the TDNKS and exact analytical results. Fig.\ \ref{Detune}(b) shows numerical results for interacting electrons (the Hartree and exchange-correlation potentials are included). In this case, Eq.\ (\ref{period}) is not applicable. The influence of two-electron excitations (difference between the TDNKS and TDKS results for the same exchange-correlation functional) is much larger than that of the exchange-correlation functional (difference between the LDA and PBE results for the same formulation), which signifies that the failure of the TDKS approach to fully describe the Rabi oscillation is not due to the exchange-correlation functional. If the resonant Rabi oscillation has the largest period, then shifts of the maxima of the TDNKS curves from $\Delta=0$ indicate that the adiabatic LDA and PBE functionals do not predict accurate Rabi periods. Note that the difference between the TDNKS and TDKS results is only significant near $\Delta=0$. For large $\Delta$ the population of two-electron excitations is small enough to be neglected, for example when $\Delta>0.6\ {\rm eV}$ it is less than 0.05, which indicates that the TDKS approach works as well as the TDNKS approach in such cases.

% FIGURE #4
%

%%%%%%%%%%%%%%%%%%%%%%%%%%%%%%%%%%%%%%%%%%%%%%%%%%
%\section{Conclusions}
%%%%%%%%%%%%%%%%%%%%%%%%%%%%%%%%%%%%%%%%%%%%%%%%%%

RT-TD-DFT equips researchers with the ability to quantify the explicit time evolution of the electronic structure in response to light-matter interaction. Because this technique continues to move into the main stream of physics and chemistry research, it is especially important to address the issue of erroneous electron density. This problem has been known for many years and has been attributed to several sources. In the present work we have shown that it results from unphysical multi-electron states, which are created through the enforcement of a fixed number of KS orbitals with fixed occupations. 

A modest modification to the existing formulation replaces the standard KS orbitals with the natural orbitals \cite{Natural2010} and updates those on the fly along with their occupations. This new TDNKS formulation improves the electron density, as demonstrated by successful simulations of Rabi oscillation on an ${\rm H_2}$ molecule subjected to an oscillatory electric field. The lowest excited state of the ${\rm H_2}$ molecule is composed exclusively of one-electron determinants for symmetry reasons. In contrast to the TDKS approach, the TDNKS approach accurately predicts a full occupation of the one-electron excited state and period of Rabi oscillation in comparison with the analytical result for the case of non-interacting electrons. Calculation of the dipole moment also shows that the TDNKS approach captures the correct electron density in contrast to the TDKS approach. While the Rabi setting makes the issue and its resolution particularly clear, the problem also exists in spin-restricted setups such as high harmonic generation and photon-induced electron ejection of molecules. This demonstrates the great potential of the proposed TDNKS approach.

%%%%%%%%%%%%%%%%%%%%%%%%%%%%%%%%%%%%%%%%%%%%%%%%%%
%\section{Acknowledgements}
%%%%%%%%%%%%%%%%%%%%%%%%%%%%%%%%%%%%%%%%%%%%%%%%%%
The authors thank Prof.\ Xiaosong Li for fruitful discussions. All calculations were
carried out using the high performance computing resources provided by the Golden Energy
Computing Organization at the Colorado School of Mines (NSF Grant No. CNS-0722415). The
research reported in this publication was supported by funding from King Abdullah
University of Science and Technology (KAUST).


\begin{thebibliography}{00}
\bibitem{RGtddftPRL1984}E. Runge and E. K. U. Gross, Density-functional theory for time-dependent systems, Phys. Rev. Lett. {\bf 52}, 997-1000 (1984).

\bibitem{TDDFT2}M. Petersilka, U. J. Gossmann, and E. K. U. Gross, Excitation energies from time-dependent density-functional theory, Phys. Rev. Lett. {\bf 76}, 1212-1215 (1996).

\bibitem{Ullrich2002}C. A. Ullrich and G. Vignale, Time-dependent current-density-functional theory for the linear response of weakly disordered systems, Phys. Rev. B {\bf 65}, 245102 (2002).

\bibitem{HeadGordon2004}A. Dreuw and M. Head-Gordon, Failure of time-dependent density functional theory for long-range charge-transfer excited states: The zincbacteriochlorin-bacteriochlorin and bacteriochlorophyll-spheroidene complexes, J. Am. Chem. Soc. {\bf 126}, 4007-4016 (2004).

\bibitem{Casida2009}M. E. Casida, H. Chermette, and D. Jacquemi, Time-dependent density-functional theory for molecules and molecular solids, J. Mol. Struct.: THEOCHEM {\bf 914}, 1-2 (2009).

\bibitem{Nakatsukasa2016}T. Nakatsukasa, K. Matsuyanagi, M. Matsuo, and K. Yabana, Time-dependent density-functional description of nuclear dynamics, Rev. Mod. Phys. {\bf 88}, 045004 (2016).

\bibitem{RTTDDFT1} G. R. Vakili-Nezhaad and G. A. Mansoori, An application of non-extensive statistical mechanics to nanosystems, J. Comput. Theor. Nanosci. {\bf 1}, 227-229 (2004).

\bibitem{Lopata2011}K. Lopata and N. Govind, Modeling fast electron dynamics with realtime time-dependent density functional theory: Application to small molecules and chromophores, J. Chem. Theory and Comput. {\bf 7}, 1344-1355 (2011).

\bibitem{Nonlinear2007}F. Wang, C. Y. Yam, and G. Chen, Time-dependent density-functional theory/localized density matrix method for dynamic hyperpolarizability, J. Chem. Phys. {\bf 126}, 244102 (2007).

\bibitem{Nonlinear2013}F. Ding, B. E. V. Kuiken, B. E. Eichinger, and X. Li, An efficient method for calculating dynamical hyperpolarizabilities using real-time timedependent density functional theory, J. Chem. Phys. {\bf 138}, 064104 (2013).

\bibitem{Xiaosong2011}C. T. Chapman, W. Liang, and X. Li, Ultrafast coherent electron-hole separation dynamics in a fullerene derivative, J. Phys. Chem. Lett. {\bf 2}, 1189-1192 (2011).

\bibitem{Cong2012}C. Wang, L. Jiang, F. Wang, X. Li, Y. Yuan, H. Xiao, H.-L. Tsai, and Y. Lu, First-principles electron dynamics control simulation of diamond under femtosecond laser pulse train irradiation, J. Phys.: Condens. Matter {\bf 24}, 275801 (2012).

\bibitem{Yabana2012}K. Yabana, T. Sugiyama, Y. Shinohara, T. Otobe, and G. F. Bertsch, Time-dependent density functional theory for strong electromagnetic field in crystalline solids, Phys. Rev. B {\bf 85}, 045134 (2012).

\bibitem{Maitra2013}J. I. Fuks, P. Elliott, A. Rubio, and N. T. Maitra, Dynamics of charge-transfer processes with time-dependent density functional theory, J. Phys. Chem. Lett. {\bf 4}, 735-739 (2013).

\bibitem{ZangTXPRA2017}X. Zang and M. T. Lusk, Twisted molecular excitons as mediators for changing the angular momentum of light, Phys. Rev. A {\bf 96}, 013819 (2017).

\bibitem{ZangTXPRB2017}X. Zang and M. T. Lusk, Angular momentum transport with twisted exciton wave packets, Phys. Rev. B {\bf 96}, 155104 (2017).

\bibitem{ZangSEPRB2017}X. Zang, S. Montangero, L. D. Carr, and M. T. Lusk, Engineering and manipulating exciton wave packets, Phys. Rev. B {\bf 95}, 195423 (2017).

\bibitem{EMD2}I. Tavernelli, U. F. R\"ohrig, and U. Rothlisberger, Molecular dynamics in electronically excited states using time-dependent density functional theory, Mol. Phys. {\bf 103}, 963-981 (2005).

\bibitem{EMD3}Y. Tateyama, N. Oyama, T. Ohno, and Y. Miyamoto, Real-time propagation time-dependent density functional theory study on the ring-opening transformation of the photoexcited crystalline benzene, J. Chem. Phys. {\bf 124}, 124507 (2006).

\bibitem{Xiaosong2007}C. M. Isborn, X. Li, and J. C. Tully, Time-dependent density functional theory Ehrenfest dynamics: Collisions between atomic oxygen and graphite clusters, J. Chem. Phys. {\bf 126}, 134307 (2007).

\bibitem{FEMD}J. L. Alonso, X. Andrade, P. Echenique, F. Falceto, D. Prada-Gracia, and A. Rubio, Efficient formalism for large-scale ab initio molecular dynamics based on time-dependent density functional theory, Phys. Rev. Lett. {\bf 101}, 096403 (2008).

\bibitem{Xiaosong2009}C. L. Moss, C. M. Isborn, and X. Li, Ehrenfest dynamics with a time-dependent density-functional-theory calculation of lifetimes and resonant widths of charge-transfer states of ${\text{Li}}^{+}$ near an aluminum cluster surface, Phys. Rev. A {\bf 80}, 024503 (2009).

\bibitem{EMD1}J. L. Alonso, A. Castro, P. Echenique, V. Polo, A. Rubio, and D. Zueco, Ab initio molecular dynamics on the electronic Boltzmann equilibrium distribution, New J. Phys. {\bf 12}, 083064 (2010).

\bibitem{Meng2010}S. Meng and E. Kaxiras, Electron and hole dynamics in dye-sensitized solar cells: Influencing factors and systematic trends, Nano Lett. {\bf 10}, 1238-1247 (2010).

\bibitem{Falke2014}S. M. Falke, C. A. Rozzi, D. Brida, M. Maiuri, M. Amato, E. Sommer, A. D. Sio, A. Rubio, G. Cerullo, E. Molinari, and C. Lienau, Coherent ultrafast charge transfer in an organic photovoltaic blend, Science {\bf 344}, 1001-1005 (2014).

\bibitem{FuksPRL2015}J. I. Fuks, K. Luo, E. D. Sandoval, and N. T. Maitra, Time-resolved spectroscopy in time-dependent density functional theory: An exact condition, Phys. Rev. Lett. {\bf 114}, 183002 (2015).

\bibitem{BauerPRL2009}M. Ruggenthaler and D. Bauer, Rabi oscillations and few-level approximations in time-dependent density functional theory, Phys. Rev. Lett. {\bf 102}, 233001 (2009).

\bibitem{FuksPRB2011}J. I. Fuks, N. Helbig, I. V. Tokatly, and A. Rubio, Nonlinear phenomena in time-dependent density-functional theory: What Rabi oscillations can teach us, Phys. Rev. B {\bf 84}, 075107 (2011).

\bibitem{FuksPRA2013}J. I. Fuks, M. Farzanehpour, I. V. Tokatly, H. Appel, S. Kurth, and A. Rubio, Time-dependent exchange-correlation functional for a Hubbard dimer: Quantifying nonadiabatic effects, Phys. Rev. A {\bf 88}, 062512 (2013).

\bibitem{BradleyJCP2014}B. F. Habenicht, N. P. Tani, M. R. Provorse, and C. M. Isborn, Two-electron Rabi oscillations in real-time time-dependent density-functional theory, J. Chem. Phys. {\bf 141}, 184112 (2014).

\bibitem{Natural2010}N. Helbig, I. V. Tokatly, and A. Rubio, Physical meaning of the natural orbitals: Analysis of exactly solvable models, Phys. Rev. A {\bf 81}, 022504 (2010).

\bibitem{Szabo1989}A. Szabo and N. S. Ostlund, {\it Modern quantum chemistry: Introduction to advanced electronic structure theory} (McGraw-Hill, New York, 1989).

\bibitem{TimePropState}C. F. Craig, W. R. Duncan, and O. V. Prezhdo, Trajectory surface hopping in the time-dependent Kohn-Sham approach for electron-nuclear dynamics, Phys. Rev. Lett. {\bf 95}, 163001 (2005).

\bibitem{Marques2012}M. A. L. Marques, N. T. Maitra, F. M. S. Nogueira, E. K. U. Gross, and A. Rubio, {\it Fundamentals of time-dependent density functional theory} (Springer, Berlin, 2012).

\bibitem{OCTOPUS}X. Andrade, D. Strubbe, U. D. Giovannini, A. H. Larsen, M. J. T. Oliveira, J. Alberdi-Rodriguez, A. Varas, I. Theophilou, N. Helbig, M. J. Verstraete, L. Stella, F. Nogueira, A. Aspuru-Guzik, A. Castro, M. A. L. Marques, and A. Rubio, Real-space grids and the Octopus code as tools for the development of new simulation approaches for electronic systems, Phys. Chem. Chem. Phys. {\bf 17}, 31371-31396 (2015).

\bibitem{Band2006}Y. B. Band, {\it Light and matter electromagnetism, optics, spectroscopy and lasers} (John Wiley and Sons, Chichester, 2006).

\end{thebibliography}
\end{document}

\title{Supplemental Material \\ Identification and Resolution of Unphysical Multi-Electron Excitations in the Real-Time Time-Dependent Kohn-Sham Formulation}

\author{Xiaoning Zang}
\affiliation{Physical Sciences and Engineering Division (PSE), King Abdullah University of Science and Technology (KAUST), Thuwal, 23955-6900, Saudi Arabia}
\author{Udo Schwingenschl\"ogl}
\affiliation{Physical Sciences and Engineering Division (PSE), King Abdullah University of Science and Technology (KAUST), Thuwal, 23955-6900, Saudi Arabia}
\author{Mark T. Lusk}
\affiliation{Department of Physics, Colorado School of Mines, Golden, CO 80401, USA}

\maketitle

%%%%%%%%%%%%%%%%%%%%%%%%%%%%%%%%%%%%%%%%%%%%%%%%%
\section{\rom{1}. Correspondence}
%%%%%%%%%%%%%%%%%%%%%%%%%%%%%%%%%%%%%%%%%%%%%%%%%

A given time-dependent one-electron reduced density may correspond to different pairs of the external/KS potential and initial wave function of the actual/KS system~\cite{Marques2012}. In applications of TD-DFT, however, a ground state is calculated first, which guarantees that there is only one choice. According to the Hohenberg-Kohn theorems, there is a one-to-one correspondence between the ground-state electron density and the ground-state wave function for both the actual and KS systems. Thus, the one-to-one correspondence between the ground-state actual and KS wave functions is established through their common electron density. When the ground-state wave function is the initial wave function the one-to-one correspondence between the actual and KS excited-state wave functions is established through the Runge-Gross theorem, more specifically through the Casida formulation of TD-DFT. The excited-state KS wave functions of the Casida formulation are, in general, not the actual wave functions, but characterize the actual excited states in the sense that their energies are good approximations to the actual excited-state energies. Thus, there is a one-to-one correspondence between the actual and KS excited states, see Fig. 1(a) in the main text. In other words, whenever the KS wave function is that of the first excited state, for example, we know that the actual system is in its first excited state. Therefore, the population of the excited-state KS wave function is identical to the population of the actual excited-state wave function.

The topic is related to v- and N-representability of an electron density in ground-state DFT~\cite{Parr1994}. The Hohenberg-Kohn theorems prove one-to-one correspondence between the v-representable density and ground-state wave function. Practically, the ground state is determined by energy minimization over all N-representable one-electron densities, since necessary and sufficient conditions for N-representability are known, but not for v-representability. An N-representable electron density minimizing the energy functional, in general, corresponds to several antisymmetric wave functions among which there is one ground-state wave function. Similarly, the choice of the initial wave function of a TD-DFT calculation is restricted to the ground-state wave function among all wave functions providing the same density. Note that in real-time TDKS simulations the initial state can be a specific excited state obtained after a ground state calculation, implying that the one-to-one correspondence is established.

%%%%%%%%%%%%%%%%%%%%%%%%%%%%%%%%%%%%%%%%%%%%%%%%%
\section{\rom{2}. Populations in the TDKS formulation}
%%%%%%%%%%%%%%%%%%%%%%%%%%%%%%%%%%%%%%%%%%%%%%%%%

The time-propagated wave function, Eq. (6) in the main text, can be written as 
\begin{equation}
\ket{\Psi}=P_{gs}\ket{\phi_i(\vec r)\bar{\phi}_i(\vec r)}+P_{1e}\frac{\ket{\phi_i(\vec r)\bar{\phi}_r(\vec r)}-\ket{\bar{\phi}_i(\vec r)\phi_r(\vec r)}}{\sqrt{2}}+P_{2e}\ket{\phi_r(\vec r)\bar{\phi}_r(\vec r)}\nonumber
\end{equation} 
with the definitions
\begin{eqnarray}
P_{gs}&=&a_{ii}^2(t)\nonumber\\
P_{1e}&=&\sqrt{2}a_{ii}(t)a_{ir}(t)\nonumber\\
P_{2e}&=&a_{ir}^2(t)\nonumber.
\end{eqnarray}
Thus, $P_{1e}=2\sqrt{P_{gs}}\sqrt{P_{2e}}$. Due to the condition $P_{gs}+P_{1e}+P_{2e}=1$, we have
\begin{equation}
P_{gs}+2\sqrt{P_{gs}}\sqrt{P_{2e}}+P_{2e}=(\sqrt{P_{gs}}+\sqrt{P_{2e}})^2=1\nonumber,
\end{equation}
which gives $\sqrt{P_{gs}}+\sqrt{P_{2e}}=1$, because all populations are real and positive. This implies Eq. (7) in the main text.

%%%%%%%%%%%%%%%%%%%%%%%%%%%%%%%%%%%%%%%%%%%%%%%%%
\section{\rom{3}. Energy conservation in the TDNKS formulation}
%%%%%%%%%%%%%%%%%%%%%%%%%%%%%%%%%%%%%%%%%%%%%%%%%

For the TDNKS formulation to be physically meaningful energy conservation must be satisfied,
\begin{equation}
\frac{dE(t)}{dt}=\braket{\partial_t \nu_{ext}(t)},
\label{Econserve}\end{equation}
where $E(t)$ is the total energy and $\braket{\partial_t \nu_{ext}(t)}$ is the expectation value of the time derivative of the external potential (the total energy is constant when the external potential is time-independent). For the M-electron KS system we have the Hamiltonian $H_{KS}=\sum_{i=1}^Mh_{KS}(\vec r_i)$ with $h_{KS}= -\frac{\hbar^2}{2m_e}\triangledown^2 +\nu_{ext}(\vec r,t)+\nu_{Hartree}[\rho](\vec r,t)+\nu_{xc}[\rho](\vec r,t)$. Using the time-dependent multi-electron wave function in real space, $\Psi(\vec {r_1},\cdots,\vec {r_M},t)$, the one-electron reduced density matrix is defined as
\begin{equation}
\rho(\vec {r_1}^\prime,\vec {r_1},t)=M\int\cdots\int\Psi^*(\vec {r_1}^\prime,\cdots,\vec {r_M},t)\Psi(\vec {r_1},\cdots,\vec {r_M},t)d^3\vec {r_2}\cdots d^3\vec {r_M}.
\label{1eDensity}\end{equation}
The time-dependent energy can be written as a functional of electron density~\cite{Parr1994},
\begin{equation}
E[\rho](t)=T_s[\rho](t)+E_{Hartree}[\rho](t)+E_{ext}[\rho](t)+E_{xc}[\rho](t),
\label{Evst}\end{equation}
where $T_s[\rho]$, $E_{Hartree}[\rho]$, $E_{ext}[\rho]$, and $E_{xc}[\rho]$ denote the kinetic, Hartree, external, and exchange-correlation energy functionals, respectively. The derivation of energy conservation is simplified by defining the energy functionals in terms of the one-electron reduced density matrix:
\begin{eqnarray}
&&T_s[\rho](t)=\sum_m^{N(t)}p_m\braket{\psi_m|-\frac{\hbar^2\triangledown^2}{2m_e}|\psi_m}\nonumber\\
&&=\int d^3\vec {r_1}\left\{M\int\cdots\int\Psi^*(\vec {r_1}^\prime,\cdots,\vec {r_M},t)\left(-\frac{\hbar^2\triangledown_1^2}{2m_e}\right)\Psi(\vec {r_1},\cdots,\vec {r_M},t)d^3\vec {r_2}\cdots d^3\vec {r_M}\right\}_{\vec {r_1}^\prime=\vec {r_1}}\nonumber\\
&&={\rm Tr}_1\left\{-\frac{\hbar^2\triangledown_1^2}{2m_e}\rho(\vec {r_1}^\prime,\vec {r_1},t)\right\},
\label{Ts}\\
&&E_{Hartree}[\rho](t)=\frac{1}{2}\int\int\frac{\rho(\vec {r_1},t)\rho(\vec {r_2},t)}{|\vec {r_1}-\vec {r_2}|}d^3\vec {r_1} d^3\vec {r_2}\nonumber\\
&&=\frac{1}{2}{\rm Tr}_1{\rm Tr}_2\left\{\frac{1}{|\vec {r_1}-\vec {r_2}|}\rho(\vec {r_1}^\prime,\vec {r_1},t)\rho(\vec {r_2}^\prime,\vec {r_2},t)\right\}
\label{EHa},\\
&&E_{ext}[\rho](t)={\rm Tr}_1\left\{\nu_{ext}(\vec {r_1},t)\rho(\vec {r_1}^\prime,\vec {r_1},t)\right\}.
\label{Eext}\end{eqnarray}
We obtain the derivative
\begin{eqnarray}
&&\frac{d T_s[\rho]}{d t}={\rm Tr}_1\left\{-\frac{\hbar^2\triangledown_1^2}{2m_e}\frac{\partial \rho(\vec {r_1}^\prime,\vec {r_1},t)}{\partial t}\right\}_1\nonumber\\
&&=\int d^3\vec {r_1}\left\{M\int\cdots\int\frac{\partial\Psi^*(\vec {r_1}^\prime,\cdots,\vec {r_M},t)}{\partial t}\left(-\frac{\hbar^2\triangledown_1^2}{2m_e}\right)\Psi(\vec {r_1},\cdots,\vec {r_M},t)d^3\vec {r_2}\cdots d^3\vec {r_M}\right\}_{\vec {r_1}^\prime=\vec {r_1}}\nonumber\\
&&+\int d^3\vec {r_1}\left\{M\int\cdots\int\Psi^*(\vec {r_1}^\prime,\cdots,\vec {r_M},t)\left(-\frac{\hbar^2\triangledown_1^2}{2m_e}\right)\frac{\partial\Psi(\vec {r_1},\cdots,\vec {r_M},t)}{\partial t}d^3\vec {r_2}\cdots d^3\vec {r_M}\right\}_{\vec {r_1}^\prime=\vec {r_1}}\nonumber\\
&&=-\frac{\mathrm{i}}{\hbar}\int d^3\vec {r_1}\left\{M\int\cdots\int\Psi^*(\vec {r_1}^\prime,\cdots,\vec {r_M},t)\left[-\frac{\hbar^2\triangledown_1^2}{2m_e},H_{KS}\right]\Psi(\vec {r_1},\cdots,\vec {r_M},t)d^3\vec {r_2}\cdots d^3\vec {r_M}\right\}_{\vec {r_1}^\prime=\vec {r_1}}\nonumber\\
&&=-\frac{\mathrm{i}}{\hbar}{\rm Tr}_1\left\{\left[(-\frac{\hbar^2\triangledown_1^2}{2m_e}),H_{KS}\right]\rho(\vec {r_1}^\prime,\vec {r_1},t)\right\}=-\frac{\mathrm{i}}{\hbar}{\rm Tr}_1\left\{\left[-\frac{\hbar^2\triangledown_1^2}{2m_e},h_{KS}(\vec {r_1})\right]\rho(\vec {r_1}^\prime,\vec {r_1},t)\right\},
\label{dTs}\end{eqnarray}
where we have employed the relations $\frac{\partial \Psi}{\partial t}=-\frac{\mathrm{i}}{\hbar}H_{KS}\Psi$, $\frac{\partial \Psi^*}{\partial t}=\frac{\mathrm{i}}{\hbar}\Psi^*H_{KS}$, and $[-\frac{\hbar^2\triangledown_1^2}{2m_e},h_{KS}(\vec {r_i})]=0$ for $i\neq1$. Similarly, we obtain
\begin{eqnarray}
&&\frac{d E_{ext}[\rho]}{d t}={\rm Tr}_1\left\{\frac{\partial \nu_{ext}(\vec {r_1},t)}{\partial t}\rho(\vec {r_1}^\prime,\vec {r_1},t)\right\}+{\rm Tr}_1\left\{\nu_{ext}(\vec {r_1},t)\frac{\partial \rho(\vec{r_1}^\prime,\vec{r_1},t)}{\partial t}\right\}\nonumber\\
&&={\rm Tr}_1\left\{\frac{\partial \nu_{ext}(\vec{r_1},t)}{\partial t}\rho(\vec{r_1}^\prime,\vec {r_1},t)\right\}-\frac{\mathrm{i}}{\hbar}{\rm Tr}_1\left\{\left[\nu_{ext}(\vec {r_1},t),h_{KS}(\vec {r_1})\right]\rho(\vec {r_1}^\prime,\vec{r_1},t)\right\}\label{dEext},\\
&&\frac{d E_{Hartree}[\rho]}{d t}=\frac{1}{2}{\rm Tr}_1{\rm Tr}_2\left\{\frac{1}{|\vec {r_1}-\vec {r_2}|}\frac{\partial \rho(\vec {r_1}^\prime,\vec {r_1},t)}{\partial t}\rho(\vec {r_2}^\prime,\vec {r_2},t)\right\}\nonumber\\
&&+\frac{1}{2}{\rm Tr}_1{\rm Tr}_2\left\{\frac{1}{|\vec {r_1}-\vec {r_2}|}\rho(\vec {r_1}^\prime,\vec {r_1},t)\frac{\partial \rho(\vec {r_2}^\prime,\vec {r_2},t)}{\partial t}\right\}\nonumber\\
&&=\frac{1}{2}{\rm Tr}_1\left\{\nu_{Hartree}(\vec {r_1})\frac{\partial \rho(\vec {r_1}^\prime,\vec {r_1},t)}{\partial t}\right\}+\frac{1}{2}{\rm Tr}_2\left\{\nu_{Hartree}(\vec {r_2})\frac{\partial \rho(\vec {r_2}^\prime,\vec {r_2},t)}{\partial t}\right\}\nonumber\\
&&={\rm Tr}_1\left\{\nu_{Hartree}(\vec {r_1})\frac{\partial \rho(\vec {r_1}^\prime,\vec {r_1},t)}{\partial t}\right\}\nonumber\\
&&=-\frac{\mathrm{i}}{\hbar}{\rm Tr}_1\left\{\left[\nu_{Hartree}(\vec {r_1}),hs(\vec {r_1})\right]\rho(\vec {r_1}^\prime,\vec {r_1},t)\right\},
\label{dEHa}\end{eqnarray}
where we have used the definition $\nu_{Hartree}(\vec {r_1})={\rm Tr}_2\left\{\frac{1}{|\vec {r_1}-\vec {r_2}|}\rho(\vec {r_2}^\prime,\vec {r_2},t)\right\}$. Note that Eqs.\ (\ref{dTs}) and (\ref{dEext}) are the Ehrenfest theorem. As an example, we examine the LDA exchange-correlation functional,
\begin{equation}
E_{xc}^{LDA}[\rho]=\int\rho(\vec {r_1},t)\epsilon_{xc}(\rho(\vec {r_1},t))d^3\vec {r_1},
\label{Exc}\end{equation}
and obtain
\begin{eqnarray}
\frac{d E_{xc}^{LDA}[\rho]}{d t}&=&\int\left(\epsilon_{xc}(\rho(\vec {r_1},t))+\rho(\vec {r_1},t)\frac{\delta\epsilon_{xc}(\rho(\vec {r_1},t))}{\delta \rho(\vec {r_1},t)}\right)\frac{\partial \rho(\vec {r_1},t)}{\partial t}d^3\vec {r_1}\nonumber\\
&=&\int\nu_{xc}^{LDA}\frac{\partial \rho(\vec {r_1},t)}{\partial t}d^3\vec {r_1}\nonumber\\
&=&-\frac{\mathrm{i}}{\hbar}{\rm Tr}_1\left\{\left[\nu_{xc}^{LDA}(\vec {r_1}),h_{KS}(\vec {r_1})\right]\rho(\vec {r_1}^\prime,\vec {r_1},t)\right\},
\label{dExc}\end{eqnarray}
with $\nu_{xc}^{LDA}=\frac{\delta E_{xc}^{LDA}[\rho]}{\delta \rho}=\epsilon_{xc}(\rho(\vec {r_1},t))+\rho(\vec {r_1},t)\frac{\delta\epsilon_{xc}(\rho(\vec {r_1},t))}{\delta \rho(\vec {r_1},t)}$.
Eq.\ (\ref{Evst}) then yields
\begin{eqnarray}
\frac{d E[\rho](t)}{d t}&=&-\frac{\mathrm{i}}{\hbar}{\rm Tr}_1\left\{\left(-\frac{\hbar^2\triangledown_1^2}{2m_e}+\nu_{ext}(\vec {r_1},t)+\nu_{Hartree}(\vec {r_1})+\nu_{xc}^{LDA}(\vec {r_1}),h_{KS}(\vec {r_1})\right)\rho(\vec {r_1}^\prime,\vec {r_1},t)\right\}\nonumber\\
&+&{\rm Tr}_1\left\{\frac{\partial \nu_{ext}(\vec {r_1},t)}{\partial t}\rho(\vec {r_1}^\prime,\vec {r_1},t)\right\}\nonumber\\
&=&-\frac{\mathrm{i}}{\hbar}{\rm Tr}_1\left\{\left[h_{KS}(\vec {r_1}),h_{KS}(\vec {r_1})\right]\rho(\vec {r_1}^\prime,\vec {r_1},t)\right\}+{\rm Tr}_1\left\{\frac{\partial \nu_{ext}(\vec {r_1},t)}{\partial t}\rho(\vec {r_1}^\prime,\vec {r_1},t)\right\}\nonumber\\
&=&{\rm Tr}_1\left\{\frac{\partial \nu_{ext}(\vec {r_1},t)}{\partial t}\rho(\vec {r_1}^\prime,\vec {r_1},t)\right\}=\braket{\partial_t\nu_{ext}(t)},
\label{dE}\end{eqnarray}
which proves Eq.\ (\ref{Econserve}). Note that this derivation is also valid for the TDKS formulation~\cite{Mundt2007}. 

%%%%%%%%%%%%%%%%%%%%%%%%%%%%%%%%%%%%%%%%%%%%%%%%%
\section{\rom{4}. TDNKS Scheme}
%%%%%%%%%%%%%%%%%%%%%%%%%%%%%%%%%%%%%%%%%%%%%%%%%

% FIGURE #2
%
\begin{figure}[ht]
\centerline{\includegraphics[width=0.5\textwidth]{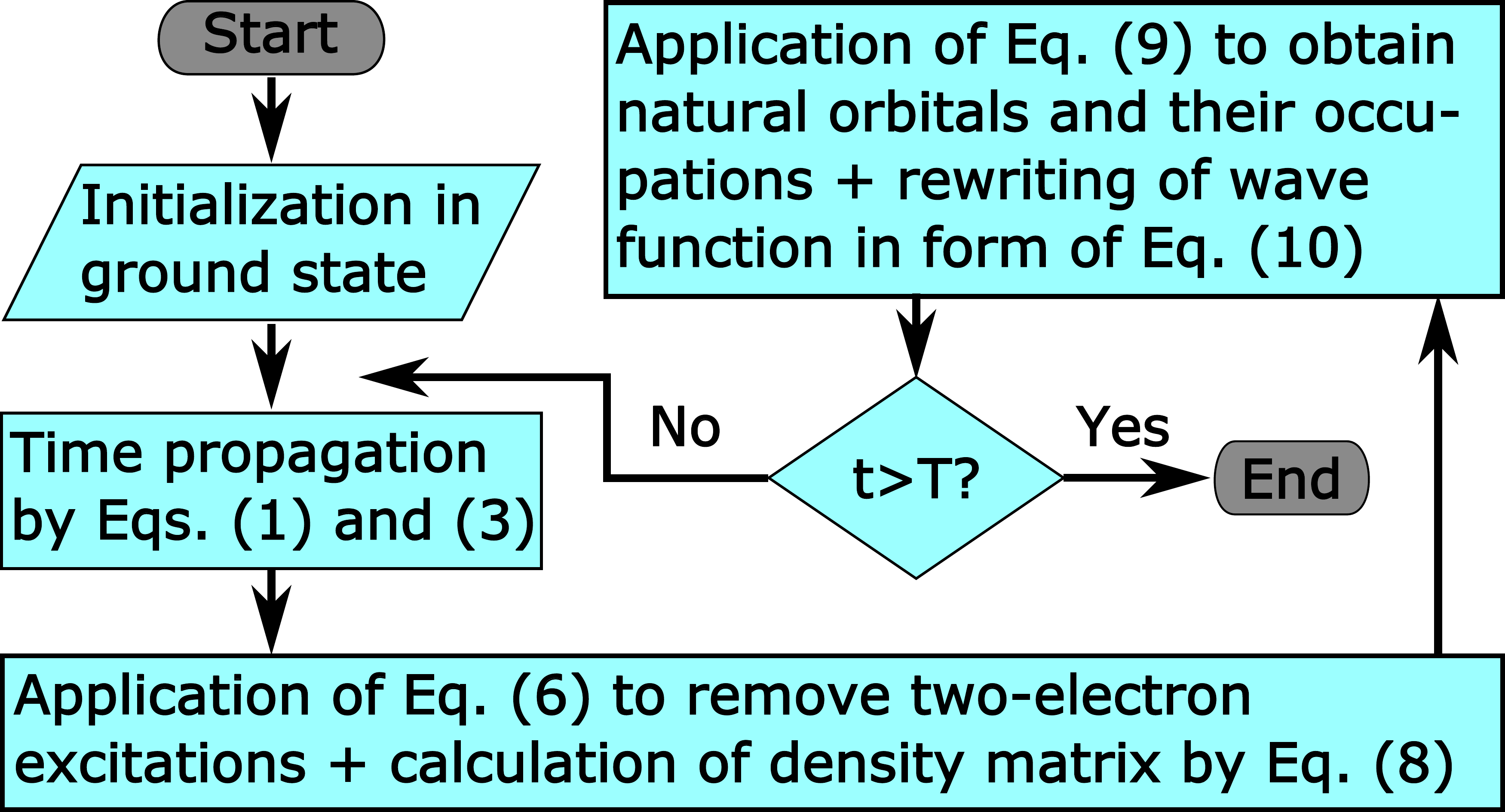}}\vspace*{-0.3cm}
\caption{Procedure of a TDNKS simulation of duration $T$.}
\label{Flowchart}
\end{figure}